\documentclass[12pt]{JHEP3}

\usepackage{multicol}
\usepackage[dvips]{epsfig,graphics}

\def\Red{}
\def\Green{}
\def\Magenta{}
\def\Blue{}
\def\Black{}

\newcommand\red[1]{{\Red#1\Black}}
\newcommand\green[1]{{\Green#1\Black}}
\newcommand\blue[1]{{\Blue#1\Black}}
\newcommand\magenta[1]{{\Magenta#1\Black}}
\newcommand{\s}[1]{\ensuremath{\widetilde{#1}}}

\newcommand{\neut}[1]{\ensuremath{\s{\chi}^0_{#1}}}
\newcommand{\charg}[1]{\ensuremath{\s{\chi}^{\pm}_{#1}}}

 \newcommand{\bmat}{\left(\begin{array}}
 \newcommand{\emat}{\end{array}\right)}

 \def\beq{\begin{equation}}
 \def\eeq{\end{equation}}
 \def\beqa{\begin{eqnarray}}
 \def\eeqa{\end{eqnarray}}

 \newlength{\wth}
 \newcommand{\fourgraphs}[4]{%
 \unitlength=1in

 \begin{picture}(5.8,4)
 \put(0,-0.2){\epsfig{file=#3, width=0.9\wth}}
 \put(2.9,-0.2){\epsfig{file=#4, width=0.9\wth}}
 \put(0,2.1){\epsfig{file=#1, width=0.9\wth}}
 \put(2.9,2.1){\epsfig{file=#2, width=0.9\wth}}
 \put(0.1,1.7){(c)}
 \put(0.1,4.){(a)}
 \put(3.0,1.7){(d)}
 \put(3.0,4.){(b)}
 \end{picture}
}

 \newcommand{\ie}{\emph{i.e.}\ }
 \newcommand{\eg}{\emph{e.g.}\ }
 \newcommand{\dl}{\ensuremath{\delta}}
 \newcommand{\dg}{\ensuremath{{}^\circ}}
 \newcommand{\mgr}{\ensuremath{m_{3/2}}}
 
 \newcommand{\agut}{\ensuremath{\alpha_\mathrm{GUT}}}
 \newcommand{\gev}{\ensuremath{\mathrm{GeV}}}
 \newcommand{\tev}{\ensuremath{\mathrm{TeV}}}
 \newcommand{\mod}[1]{\ensuremath{|#1|}}
 \newcommand{\tb}{\ensuremath{\tan\beta}}
 \newcommand{\tanb}{\tb}
 \newcommand{\vv}{\ensuremath{\vec{v}}}
 \newcommand{\thet}{\ensuremath{\theta}}
 \newcommand{\ph}{\ensuremath{\phi}}
\newcommand{\mc}[1]{\ensuremath{\mathcal{#1}}}
 \newcommand{\minuit}{\textsc{Minuit}}
 
 \newcommand{\softs}{{\tt \small SOFTSUSY1.8.4}}

\newcommand{\lsim}{\;\raise0.3ex\hbox{$<$\kern-0.75em\raise-1.1ex\hbox{$\sim$}}\
;}
\newcommand{\gsim}{\;\raise0.3ex\hbox{$>$\kern-0.75em\raise-1.1ex\hbox{$\sim$}}\
;}

\newcommand{\susy}[1]{\ensuremath{\tilde{#1}}}

\title{Genetic Algorithms and Experimental Discrimination of SUSY Models}

\author{B.C. Allanach \\
LAPTH, 9 chemin Bellevue, BP110, Annecy 74941, France\\
E-mail: \email{benjamin.allanach@cern.ch}}
\author{D. Grellscheid \\
Physikalisches Institut der Universit\"at Bonn,
Nussallee 12, 53115 Bonn, Germany\\
E-mail: \email{grelli@th.physik.uni-bonn.de}}
\author{F. Quevedo\\
DAMTP, Centre for Mathematical Sciences
               University of Cambridge,
               Cambridge, CB3 0WA, United Kingdom\\
E-mail: \email{f.quevedo@damtp.cam.ac.uk}}

\abstract{We introduce genetic algorithms as a means  to
estimate the accuracy required to 
discriminate among different models using
experimental observables. 
We exemplify the technique in the context of the minimal supersymmetric
standard model. 
If supersymmetric particles are discovered, 
models of supersymmetry breaking will be fit to the observed spectrum
and it is beneficial to ask beforehand: what accuracy is required 
to always allow the discrimination of two particular
models and which are the most important masses to observe?
Each model predicts a bounded patch in the space of observables once unknown
parameters are scanned over.
The questions can be answered by minimising a ``distance'' measure
between the two hypersurfaces. 
We construct a distance measure that scales 
like a constant fraction of an observable.

Genetic algorithms, including  concepts such as  natural selection,
fitness and  mutations,   provide a solution to the minimisation problem.
We illustrate the efficiency of the method by comparing three
different classes of string models for which the above questions could not be
answered with previous techniques. 
The required accuracy is in the range accessible to the Large Hadron
Collider
(LHC) 
when combined with a future linear collider (LC) facility.
The technique presented here can be applied to more general classes of models
or observables.
}


\preprint{DAMTP-2003-142\\  hep-ph/0406277}

\begin{document}

\section{Introduction}

Genetic algorithms (GAs)~\cite{goldberg}  have found a plethora of applications
in different 
scientific disciplines.
They were
first studied in the 1950s
when an ingenious realisation of the natural selection
mechanism that determines the evolution of biological systems was
implemented in a concrete mathematical algorithm.
 Their novelty lies in the application of biological
ideas from evolution theory to a wide range of problems in which some
measure exists that can be equated to the \emph{fitness} of a
particular solution.
Subsequently, especially since the establishment of the
mathematical foundations of GAs, they have been applied very
successfully 
to a wide range of problems, from straightforward  extremisation to
others more intractable to traditional methods, such as timetable
scheduling, resource allocation, real time process control or design of
efficient machines. 

The basic idea of the algorithm is very simple. Given a set of points 
in which a quantity has to be optimised, the algorithm describes a
well defined procedure to select the fittest of the points, to combine
their characteristics to produce offspring which will statistically be
closer to the optimal value.
The algorithm includes mutations and
other features present in natural evolution.

To the best of our knowledge these algorithms have only been used once
in theoretical high energy physics~\cite{egeg}. 
We propose a concrete application of these
techniques in order to discriminate models beyond the Standard Model.
We will use SUSY models and sparticle masses as examples, but really any
classes of models and observables could be used.
In particular, assuming that several sparticle masses
will be measured at present and future colliders, we can ask the
questions \cite{agq}: (a) what accuracy on
sparticle 
mass measurements will be required to guarantee discrimination of 
supersymmetry (SUSY) breaking models? 
(b) Which are the most important mass variables to
measure? Even though we discuss discrimination of particular models, 
the questions are ambitious because in order to {\em guarantee}
discrimination, we must scan over all free parameters in the models being
considered. This is in contrast to more experimentally based studies (eg
Ref.~\cite{study1}) where the parameters of one  model are {\em
  fixed} 
and it is shown that {\em at that particular parameter point},
particular Large Hadron Collider (LHC) observables can discriminate against 
a certain string-inspired model.

Once one has accurate information on sparticle masses, 
one could potentially
 try to deduce the SUSY breaking terms at the electroweak scale and
evolve them up to a higher scale to see if they unify~\cite{zerwas}. Because
the phenomenologically parameterised minimal supersymmetric standard model
(MSSM) contains many free parameters,
one can only obtain accurate information on running SUSY-breaking parameters
from the pole parameters if all of the sparticles (and their mixings) 
are measured. 
This may be difficult to achieve in practice unless {\em all}
of the sparticles are
light enough to be produced and measured 
in a future linear collider facility~\cite{lc}
(LC). Another approach advocated~\cite{gordy} takes inclusive hadron collider
and indirect signatures in order to discriminate particular model points.

In a previous article
\cite{agq} we addressed the questions (a) and (b), comparing three
well defined supersymmetric models motivated from string theory.
We tried to find
projections onto 
2-dimensional sparticle mass-ratio space in which the SUSY breaking scenarios
were completely disjoint and identified ``by eye''. Mass ratios were used to 
eliminate dependence on an overall mass scale. The strategy worked for
simple cases with a very  small 
number of parameters (\eg comparing the dilaton-dominated scenario in
 three different classes of string
models~\cite{agq,aaikq,aqothers}). Interesting results were obtained for this
case, in which combined information from both LHC and LC collider
experiments would be needed to differentiate the models given the level
of accuracy  required on the experimentally measured values of the
ratios of sparticle masses. 
However, departing from dilaton domination meant that more free
parameters were introduced,
 no disjoint projections were found and the models could not be
 distinguished after knowing  the sparticle masses.
Indeed, the procedure followed 
was 
unsystematic, used a limited amount of information about sparticle
masses and was somewhat limited in scope. In this article, we 
develop a systematic algorithm based on GAs that promises to 
address the weaknesses of the previous approach. 

We begin by describing the general nature of the problem, briefly explaining
why standard minimisation methods are not suitable.
We then describe the basics
of GAs, assuming no previous knowledge.
We then apply GAs to discriminate among the three
aforementioned different scenarios motivated by low-energy string models. We
conclude with a general discussion of our results including possible future
applications of the method.

\section{Formalism}
Let us consider a supersymmetric model $n$ derived from a fundamental
theory at a large scale $M_X$. Once supersymmetry is broken, soft
breaking terms will be induced, which can be parametrised by a set
of $N_n$  parameters $\{z_i\}_{i=1}^{N_n}$. The soft breaking terms,
corresponding to  
  gaugino masses
$M(z_i)$,
scalar masses $m(z_i)$ and trilinear scalar couplings $A(z_i)$, can
all be seen as functions of these parameters. Here
we have omitted the indices on $M,m,A$. More concretely, the
parameters $z_i$ could be identified for instance with typical goldstino angles
appearing in string models, as well as  the gravitino mass
$m_{3/2}$ and the ratio of MSSM Higgs fields $\tan\beta$.

In order to compare this to direct experimental observables there is a
well defined procedure to follow.
A theoretical boundary
condition upon 
SUSY breaking masses is applied at the scale $\mu=M_X$. Empirical boundary
conditions on Standard Model gauge couplings, particle masses and mixings are
applied at the 
electroweak scale $\mu=M_Z$. 
The MSSM RGEs consist of many
coupled non-linear first-order homogeneous ordinary differential equations,
with respect to renormalisation scale $\mu$. 
The calculation of the MSSM spectrum involves solving these differential
equations while simultaneously satisfying the two boundary
conditions. Radiative corrections must be 
added in order to obtain pole masses and mixing parameters for the
sparticles and to set the Yukawa and gauge parameters from data. 
We use \softs~\cite{softsusy}, a program which is designed to solve this
problem\footnote{Several other publicly-available tools~\protect\cite{tools}
  exist to solve this problem also.}. 

Therefore we are usually presented with the problem of comparing
two different spaces of parameters.
The first is the space of free model 
parameters at the high scale, which we shall refer to as
$\mathcal{I}=\{z_i\}$. Each model $n$ under 
consideration will have its own input space $\mathcal{I}_n$. The number of its
dimensions $N_n$
 is determined by the number of free parameters in the
model.
To make our analysis technically 
feasible, this should be a
small number, typically smaller than 6--8. Each point
in $\mathcal{I}_n$ then corresponds to one fixed choice of high-scale input
parameter values for model $n$. 
The sets of parameters in each $\mathcal{I}_n$ may or may not be the same
since we are
talking about completely separate inputs for two separate scenarios.

The second space, $\mathcal{M}$, is the space of physical
measurements at the electroweak scale. There is only one
unique $\mathcal{M}$, since all of the models that we consider 
describe MSSM observables.  Its dimensionality $D$ equals the number of
low-scale observables under consideration. We take typical values that
are as large as 20--30 (i.e. most sparticle masses). Unlike with the input
parameters, however, it is possible to take far larger numbers of
observables into account without a significant increase in the
complexity of the problem.
Each point in $\mathcal{M}$  denotes the allocation of one fixed
value for every observable.

Each model $n$ also specifies a set of renormalisation group equations
(often this may be the set of standard MSSM RGEs), through which
each point in $\mathcal{I}_n$ 
can potentially be mapped onto a point in $\mathcal{M}$ 
(see figure \ref{pascosfig}).
We have to say potentially, since it is not guaranteed that all
possible input points in $\mathcal{I}_n$ will actually generate a
physical result when run through the RGEs, as they might lead \eg to a
point without the correct radiative electroweak symmetry breaking. 

A scan over all $N_n$ parameters in
$\mathcal{I}_n$ will now build up an  $N$-dimensional hypersurface
in $\mathcal{M}$, made up of one point each for every input point for which
the RGE 
running was successful.
We will call this hypersurface the \emph{footprint} of the high-scale
model under consideration. 
It is only at the level of the space $\mathcal{M}$, that we can impose our
final constraints: only such points that do not violate experimental 
bounds are considered to be a part of the footprint. All other points,
\eg with too light neutralinos or charginos,
will be discarded. The set of criteria at this level depends on some
overall assumptions about the investigation. One example for this is
the question of $R$-parity conservation. If we assume that $R$-parity
is indeed conserved, the cosmological  requirement of a neutral LSP
would also be used to discard points that show a charged LSP. 

A schematic of the parameter spaces 
is displayed in figure \ref{pascosfig} for the case of the comparison of two
scenarios of SUSY breaking. 
$\mathcal{I}_1$ and $\mathcal{I}_2$ are input parameter spaces for 
two different SUSY breaking models.
Each point within $\mathcal{I}_n$ corresponds to one set of high-scale
parameters for model $n$, serving as input to that model's RGEs. 
They uniquely map each input
point onto a point in $\mathcal{M}$, the measurement space. Scanning over
$\mathcal{I}_n$ point by point builds up the ``footprint'' of model $n$ in
$\mathcal{M}$.
The closest approach of the two footprints is indicated by $\vec{v}$ and
constitutes the important discriminating variable.
In practice, $\mathcal{I}_n$ will have a finite volume, since we will apply 
an upper bound upon sparticle masses in order to avoid large fine tuning in the
Higgs potential parameters~\cite{fineTuning}.
We will choose the dimensions of $\mathcal{M}$ such that the footprints 
also have a finite volume.
\FIGURE{
\epsfig{file=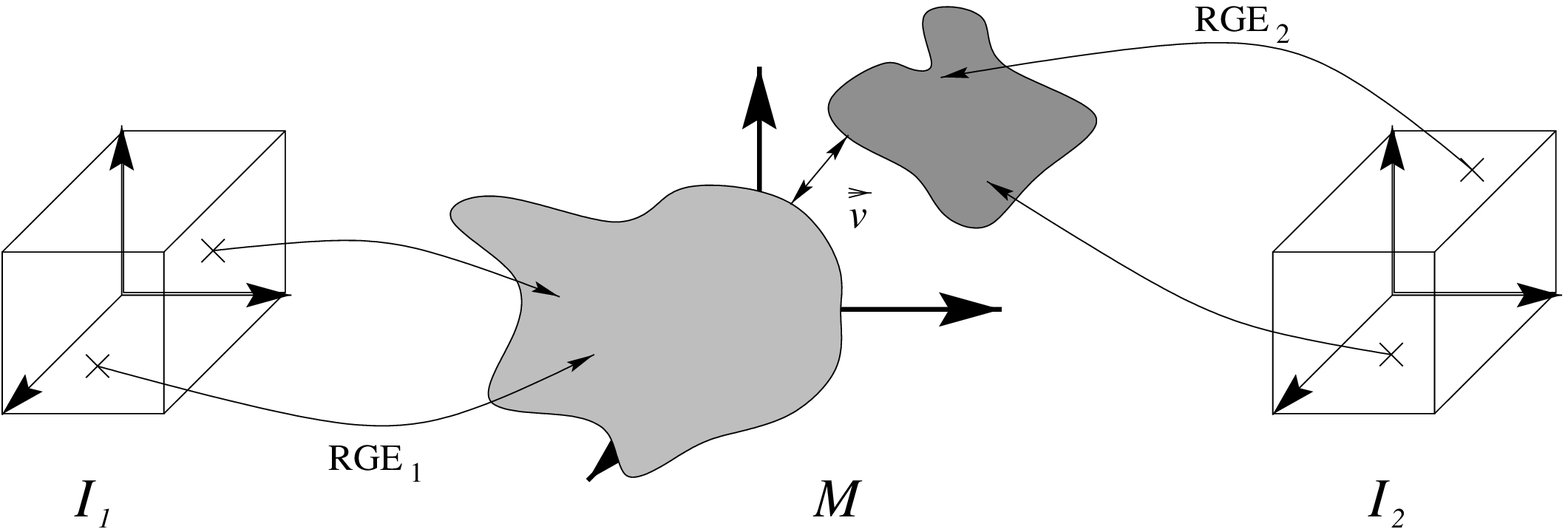,width=0.8\columnwidth}
\caption{Generating footprints from high scale parameter
  scans. 
}\label{pascosfig}}

The last
removal of points, together with the unsuccessful runs of the RGEs, implies a
``back reaction'' \label{backreaction}(see fig.~\ref{backreactionfig})
onto the high scale model spaces $\mathcal{I}_m$, by 
ruling out some groups of input points that do not lead to acceptable 
models. 
It is important to note that it is not possible to decide
on the viability of an input point in \mc{I}${}_m$ before an attempt
has actually been made at calculating the corresponding output point
in \mc{M}.
\FIGURE[]{
\epsfig{file=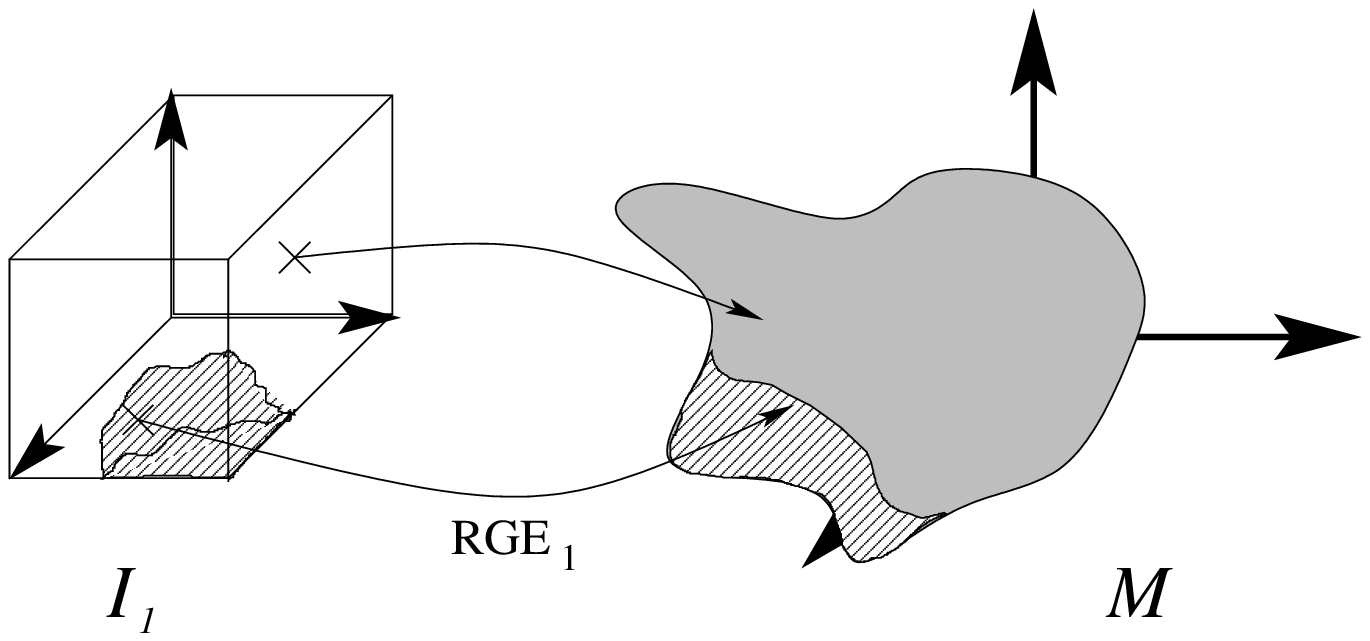,width=0.8\columnwidth}
\caption{Back reaction of non-physicality constraints onto the input
  parameters. Any point in the measurement space \mc{M} that is
  found to violate experimental bounds will not be part of the
  footprint. This in turn implies that the input point from which it
  was created is not physical. Any experimental boundaries
  in \mc{M} will thereby lead to transformed boundary lines in
  \mc{I}, which delimit a region of valid input points.
}
\label{backreactionfig}
}

Different models will have different
footprints, some of which may be disjoint, while 
others may overlap. 
However, as long as the footprints' hypersurfaces 
are of much lower dimensionality
than \mc{M}, as is usually the case, it will be quite unlikely
that there will be any overlap between the prints. 
As soon as it is established that the two prints are
disjoint, it is  possible to conclude that the two models can in
principle be distinguished experimentally, as long as a certain
 measurement accuracy can be achieved. Fig.~\ref{pascosfig} shows that this 
can be established by finding that the smallest vector $\vec{v}$ still has a
size greater than zero.

\subsection{Distance Measure}
In our previous paper \cite{agq}, we looked for the closest approach of models
in
the space spanned by dimensionless mass ratios, where $\vec{v}$ was
well defined as the smallest Euclidean distance possible. 
To make a more general statement possible, we now want to plot the model
footprints using the masses directly. The smallest Euclidean distance
is  not such a suitable measure anymore, since all calculated sparticle
masses are roughly proportional to the input value \mgr, and the
resulting vector $\vec{v}$ would always be the one closest to the
origin. 

Instead, let us look at \textit{relative distance}. In the
one-dimensional case we define this relative distance of two points 
$A$ and $B$ along one dimension to
be 
\beq\label{reldistOneDim}
\delta = \frac{|a-b|}{a+b}\,;\qquad a,b>0\,.
\eeq
This automatically guarantees $\delta \in [0,1)$, and the minimum
  value of $\delta$, if found,  can be
  seen as the relative measurement accuracy required to definitely distinguish
  the points $A$ and $B$. A distance measure such as this one scales as a
  constant fraction if one increases both $a$ and $b$ in the same ratio. 
  It is therefore useful because it gives a reasonable relative weight to the
  different variables one wishes to include in the distance measure. 
We imagine that $A$ and $B$ are the closest pair of points that can be
  predicted each by a particular different model.
Supposing one measured the observable to have value $a$. If the fractional
  experimental uncertainty (to some chosen confidence level, for example
  2$\sigma$) is smaller than $\delta$ the two points are obviously resolved by
  the   measurement, therefore the models are discriminated. Therefore
  $\delta$ is a measure of the level of discrimination needed.

We now extend this interpretation to multiple dimensions.
Let $A$ and $B$ be represented by $\vec{a}$ and $\vec{b}$:
\beq
\Delta = \frac{|\vec{a}-\vec{b}|}{|\vec{a}+\vec{b}|}
=\sqrt{\frac{(a_1-b_1)^2+\cdots+(a_D-b_D)^2}{(a_1+b_1)^2+\cdots+(a_D+b_D)^2}}.
\label{dist-measure}
\eeq
\FIGURE[]{%
\epsfig{file=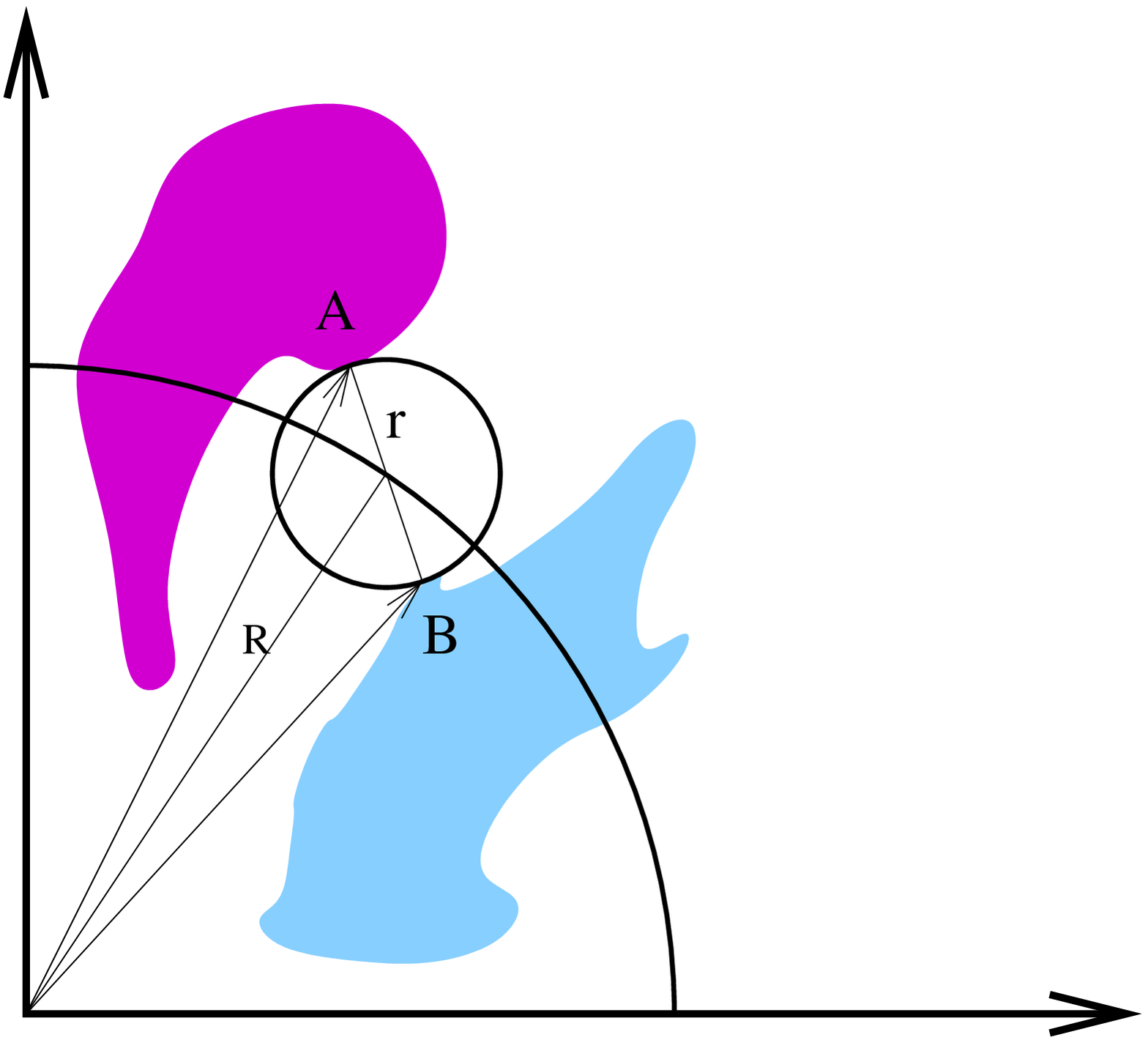,width=0.6\columnwidth}
\caption{Geometrical measure of distance. The dark and light blobs represent 
the footprints of two different models to be distinguished in the measurement
space \mc{M}, two dimensions of which are depicted.}
\label{spherepic}
}
$\Delta$ is the direct extension of
eq.~\ref{reldistOneDim} to more than one dimension, with a 
geometric interpretation $\Delta = r/R$ (see fig.~\ref{spherepic}): Let us
introduce $M$ as the midpoint of $A$ and $B$. Then $R$ is
the radius of the hypersphere around the origin which passes through
$M$, and $r$ 
is the radius of the hypersphere around $M$ with the diameter $AB$.
This property makes
the $\Delta$-measure  invariant under rotations of the
coordinate system. 
$\Delta$ gives us only a rough idea of the relative accuracy one needs to
separate the models. 
The precise meaning of $\Delta$ in terms of measurements upon
observables depends upon the details of the particular case under
study (for example, whether the footprints are convex or not, or aligned with
an axis). 
At point $B$, measuring the combination of observables in the
direction of $\vec{v}$, where 
$\vec{v}=\vec{a}-\vec{b}$, to a precision better
than $\Delta$ guarantees separation of the two points. 
Other measures sharing some of these properties could be conceived.

\subsection{Function minimisation}\label{blackbox}
We may view the
``length'' of \vv\ to be a function of two sets of input parameters.
For our example case this might be
\beq\label{d-is-fn-of-i}
\mod{\vv}=f(I_1,I_2)\equiv f\big((z_i)_{1},(z_i)_{2}\big)\, ,
\eeq
\ie we can express our problem in terms of  a scalar function of the free
parameters whose function value is to be minimised. The internal
workings of the function 
\beq\label{dvec-is-avec-minus-bvec}
f(I_1,I_2)=
\frac{\left|\vec{A}(I_1)-\vec{B}(I_2)\right|}%
{\left|\vec{A}(I_1)+\vec{B}(I_2)\right|}
\eeq
are irrelevant as far as  the search algorithms are concerned.

Now that we have reduced the problem to the maximisation of a
function, one might 
expect the application of standard techniques to provide the answer.
However, these techniques had technical problems that rendered them
insufficient to solve the problem. Performing a scan in every input parameter
would of course work, provided a fine enough scan was done. However,
there are too many input parameters for such a scan to be completed in a
reasonable amount of time. Maximisers that calculate a derivative of the
function such as \minuit~\cite{minuit} in order to implement a ``hill
climbing'' algorithm also fail. The reason for this is the back reaction
displayed in Fig.~\ref{backreactionfig}. There are many regions which cannot
be predicted in advance, where the derivative does not exist because the
region of parameter space is unphysical and no spectrum can be calculated. 
Efforts to get around this problem by assigning a ``penalty'' factor to
$\Delta$ in unphysical parts of parameter space 
failed because the maximiser calculated a spurious derivative and 
either oscillated between physical and unphysical parts of parameter space, or
got ``stuck'' in unphysical space.

\section{Genetic Algorithms}
One powerful set of tools does not suffer from any of the drawbacks
mentioned above
that made the  deterministic minimisation search by \minuit\ so
ineffective. These are \emph{genetic algorithms (GAs)}. 
In the following, at first we 
introduce GAs~\cite{goldberg} using an example 
problem, and briefly show the mathematical background behind
their success. 

\subsection{Overview}\label{ga-example}
Genetic Algorithms differ in several points from other more deterministic
 methods:  
\begin{itemize}
\item They simultaneously work on populations of solutions, rather than
tracing the  progress of one point through the problem space. This
gives them the advantage of checking many regions of the parameter
space at the same time, lowering the possibility that a global optimum
might get missed in the search.
\item They only use payoff information directly associated with each
investigated  point. No outside knowledge such as the local gradient
behaviour around  the point is necessary. For our problem this is one
of the main advantages compared to \minuit, where the calculation of
the local gradient takes a large effort in computing time. It also
makes the GA robust against points that are undefined.
\item They have a built-in mix of stochastic elements applied under
  deterministic rules, which improves their behaviour in
problems with  many local extrema, without the serious performance
  loss that a purely random search would bring.
\end{itemize}
All the power of genetic algorithms lies in the repeated application of
three basic operations onto successive generations of points in the
problem space. These are
\begin{enumerate}
\item \emph{Selection},
\item \emph{Crossover} and
\item \emph{Mutation}.
\end{enumerate}
In the following, a simple example shall illustrate their operation. 

\subsection{An Example}
As a simple problem, let us consider the maximisation of
$f(x)=x^2$ on the integer interval $x \in [0,31]$ (our example
follows the discussion in \cite{goldberg}). 
A simple analytic problem like the given one can of course be solved
much more efficiently by straightforward hill climbing algorithms. The
strength of GAs only really shows in problems that are generally hard
for deterministic optimisers. For the purpose of an introduction into
the mechanisms of a GA this will be sufficient, though.

\subsubsection{Encoding}
We need to encode our problem parameter $x$ into a string, the
\emph{chromosome}, on which
the GA can then operate. One frequent choice is a straightforward 
binary encoding, where
$x=1$ codes as \texttt{00001} and $x=31$ as \texttt{11111}. In
problems with more than one input parameter, the chromosome can simply
be formed by concatenating each parameter's string.
For the example presented here, we illustrate with binary encoding, which is
particularly easy to follow. Later, however, we find  that a real 
valued encoding is
more useful to solve our problem.

\subsubsection{Initial Population}
After this initial design decision, 
the first real step in the running of a GA is the creation of an
initial population with a fixed number of individuals
$i=1,\,\ldots,\,N$ (table~\ref{inipop}).

\TABULAR{|c|c|c|c|c|}{%
\hline
$i$ & Genotype & Phenotype $x_i$ & Fitness $f_i=f(x_i)$ & $f_i / \sum{f_i}$\\
\hline
1 & \texttt{\red{01101}}  & 13 & 169 & 0.14\\
2 & \texttt{\green{11000}} & 24 & 576 & 0.49\\
3 & \texttt{\magenta{01000}}&  8 &  64 & 0.06\\
4 & \texttt{\blue{10011}} & 19 & 361 & 0.31\\
\hline
}
{Randomly generated starting population. $i$ labels each
  individual. Their genotypes (chromosomes) are assigned randomly, and are
  translated into the phenotype value according to  the chosen encoding
  method. The fitness is then calculated with the function that we
  want to maximise: $f(x)=x^2$. The last column shows the fitness
  values normalised to 1. The colours shall make it easier to
  follow the propagation of these genes throughout the subsequent
  generations in this example. \label{inipop}}

 To make this example tractable, we use a population size of only
 four. Real applications 
regularly use populations with 50--100 individuals or more. They would
also usually have  larger chromosome sizes.

Each individual in this first population is given a randomly
generated chromosome which then represents its \emph{genotype}. Note
that this random assignment of genotypes only happens in the first
generation, but not in the subsequent ones.  
According to the
encoding we have chosen, each  chromosome implies for its owner  a
\emph{phenotype} $x_i$. The fitness of each individual is in turn a
function of 
this phenotype. This fitness value must be positive definite, and the
choice of fitness function must be such that the
problem's best solutions should be the ones with the highest fitness.
Since our problem is a maximisation with positive function values
only, we can take 
$f(x)$ directly as the fitness function.  
The last column in table \ref{inipop} shows the normalised fitness,
which will be used shortly.

To solve our optimisation  problem we now need rules that tell us how
to obtain the following generation from the present one. These rules of
course should also guarantee some improvement in the solutions over successive
generations.

\subsubsection{Selection}\label{ex-sel}
The first operator to be applied is \emph{selection}. It should ensure
that individuals with higher fitness will have a larger chance of
contributing offspring to the next generation. 
Several different selection operators can be used, here we will use one
that tries to model the selection we find in natural evolution.
The child
population in a basic GA is fixed to have the same number of individuals as the
parent population, so we need to repeat the selection of two
individuals who will 
act as parents for two children, until we have picked $N/2$ breeding pairs.
In this procedure, an individual of average fitness should be
selected about once, while individuals with higher/lower fitness
should be selected more/less frequently. 

One selection method that achieves this is \emph{roulette wheel
selection}. A visual interpretation is the following: Fix a needle
onto the centre of a pie chart of all individuals,  where the size of each
sector is proportional to that individual's fitness (see
fig.~\ref{piechart}).
\FIGURE[]{
\epsfig{file=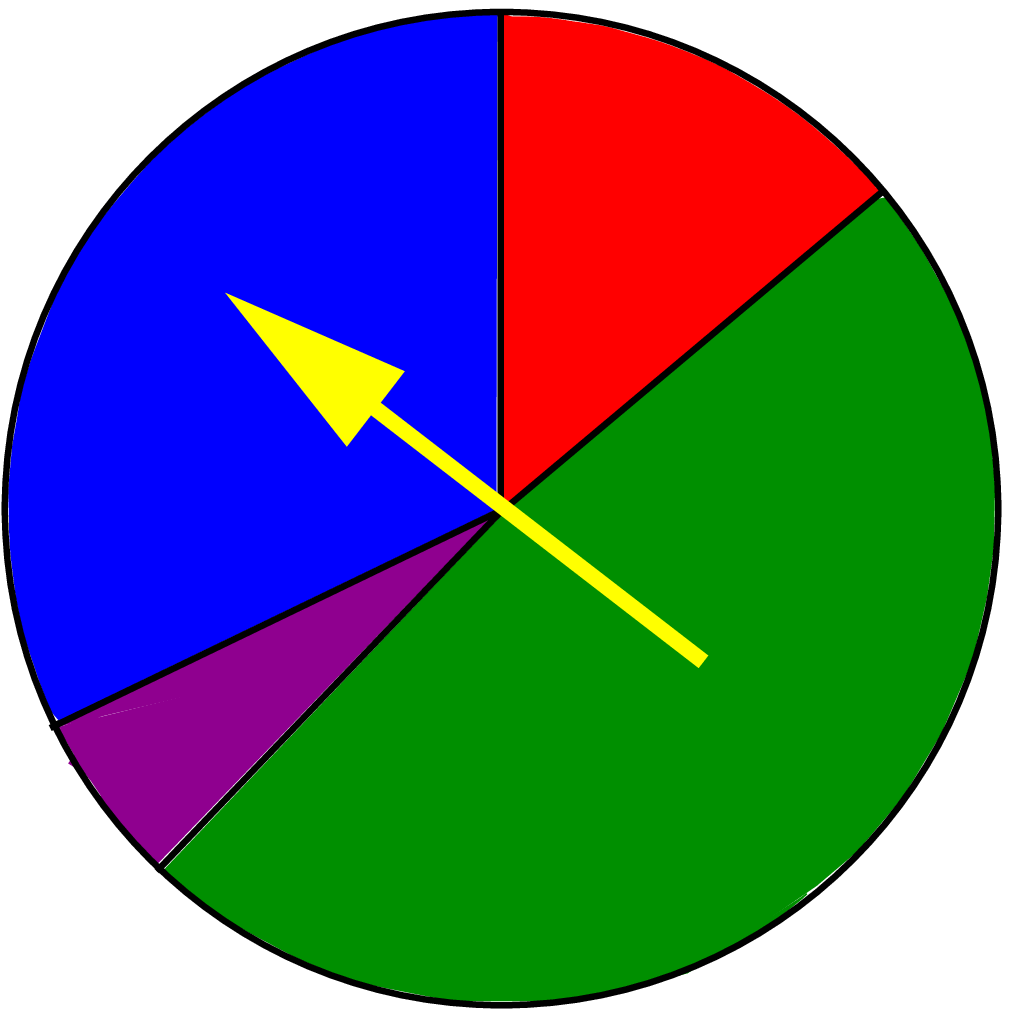,width=0.3\columnwidth}
\caption{Roulette wheel selection.}
\label{piechart}
}
Each individual in the parent
  population (table~\ref{inipop}) is assigned a
  sector of the circle, with the distended 
angle in proportion to its fitness. Now the
  needle, 
  which is assumed to end up in any position with uniform probability, 
  is turned $N$ times. Each time, the pointed-to individual is
  selected into the breeding pool, where they are paired
  up. Obviously, individuals with larger fitness will turn up in the
  breeding pool more frequently. A
  self-pairing is possible and permitted.
Since fitter individuals will
distend a larger angle on the pie chart, they have a larger chance of being
selected. 
To continue the example, let us assume that in the four spins of the
wheel the following parents were
chosen: 1, 2, 2 and 4. Note that individual 2 was chosen twice, while
3 does not 
appear at all. This corresponds nicely with the respective fitnesses. 

\subsubsection{Crossover}
The \emph{crossover} operation is necessary
to obtain a child generation that is genetically different from the
parents.
The  parents that were selected in the previous step are paired up randomly
(${ 1}\longleftrightarrow {\green{2}}$ and ${\green{2}}\longleftrightarrow 
{\blue{4}}$) and a crossover site $s$ within the chromosome is randomly
chosen for each pair. Let us assume that this was $s_1=4$ and
$s_2=2$ for the two breeding pairs respectively.

The chromosomes of both parents are cut after that position (see
table \ref{xover}),
and the ends are exchanged  to form the chromosomes for the two children.
Table \ref{childpop} shows the new generation.
\TABLE[]{
\begin{tabular}{|c|c|c|c|c|c|}
\hline
\multicolumn{3}{|c|}{Breeding pair 1} &
\multicolumn{3}{|c|}{Breeding pair 2}\\
\hline
$i$ & before & after &
$i$ & before & after \\
\hline
1 & \texttt{ \red{0110}|\red{1} } & \texttt{{ \red{0110}}|{\green{0}} } &
2 & \texttt{\green{11}|000} & \texttt{{\green{11}}|\blue{011}}\\
2 & \texttt{\green{1100}|\green{0}} & \texttt{{\green{1100}}|{\red{1}}}&
4 & \texttt{\blue{10}|011} & \texttt{{\blue{10}}|{\green{000}}}\\
\hline
\end{tabular}
\caption{\label{xover}Crossover operation on the selected parents. The
  crossover 
site was randomly chosen to be $4$ and $2$ respectively. The parts of
the chromosomes after the crossover site get swapped to create two new
chromosomes which constitute the genotypes for the two children.}
}
\TABULAR{|c|c|c|c|c|}{
\hline
$i$ & Genotype & Phenotype $x_i$ & Fitness $f_i=f(x_i)$ & $f_i / \sum{f_i}$\\
\hline\hline
5 & \texttt{{\red{0110}}{\green{0}}} & 12 & 144 & 0.08\\
6 & \texttt{{\green{1100}}{\red{1}}} & 25 & 625 & 0.36\\
7 & \texttt{{\green{11}}{\blue{011}}} & 27 & 729 & 0.42\\
8 & \texttt{{\blue{10}}{\green{000}}} & 16 & 256 & 0.14\\
\hline
}
{First child generation after selection and crossover. In
  comparison with table~\ref{inipop}, the best solution has gone up
  from a fitness of $(\max f_i)=576$ to $(\max f_i)=729$, the average
  fitness has gone up 
  from $\bar{f_i}=292.5$ to $\bar{f_i}=438.5$. This rapid increase of
  fitness over the very first few generations is a common feature of
  GAs. \label{childpop}} 
In principle, one could now go back to section \ref{ex-sel}, taking the
child generation as the new parents, and start selecting the next
breeding pairs. This ignores one danger: if one position in the
chromosomes is set to the same value in all the individuals by
accident, the crossover operator will not change that fact and one
whole subset of the problem space will not get visited by the GA. 
The final operator solves this problem.

\subsubsection{Mutation}
With a small probability
(0.01 -- 0.001) every bit in every chromosome is flipped from \texttt{1} to
\texttt{0} or
reverse. This operator is usually applied to all chromosomes after
crossover, before the fitnesses of the new generation are
evaluated. Over the span of several generations then, even a stagnated
chromosome position can become reactivated by mutation.

Even after only one generation one can observe the effects that make
GAs work. Crossover of individuals 2 and 4 has produced a child of new
higher fitness, already quite close to the theoretical optimum. Also
notable is the increase in average fitness from 292.5 to 438.5. This
is a general feature in GAs, the maximum fitness already approaches the
optimal value within the first few generations, the average fitness is
not far behind.

\subsection{Schemata make GAs work}
The theoretical concept behind the success of GAs is the concept of
patterns or \emph{schemata} within the chromosomes
\cite{goldberg}. Rather than operating 
on only $N$ individuals in each generation, a GA works with a much higher
number of schemata that partly match the actual chromosomes. 

A chromosome like \texttt{10110} matches $2^5$ schemata, such as
\texttt{**11*}, \texttt{***10} or \texttt{1*1*0}, where \texttt{*} stands
as a wild card for either \texttt{1} or \texttt{0}. Since fit
chromosomes are handed down to the next generation more often than
unfit ones, the number of copies $n_S$ of a certain schema $S$ associated with
fit chromosomes will increase from one generation to the next:

\beq\label{GAfund1}
n_S(t+1) = n_S(t) \cdot \frac{\bar{f}(S)}{\bar{f}_\mathit{total}},
\eeq
 where $\bar{f}(S)$ is the average fitness of all individuals whose
 chromosomes match schema $S$, and $\bar{f}_\mathit{total}$ is the
 average fitness of all individuals. If we assume  that a certain
 schema approximately
gives all matching chromosomes a constant fitness advantage $c$
 over the average  
\beq
\bar{f}(S) \equiv (1+c)\cdot\bar{f}_\mathit{total},
\eeq
 we get an exponential growth in the number of this schema from one
generation to the next:
\beq
\quad n_S(t) = n_S(0) \cdot (1+c)^t.
\eeq

Equation \ref{GAfund1} needs to be corrected for the effects that
crossover and mutation may have. To do this we need to define two
measures on schemata:
\begin{itemize}
\item The \emph{defining length} \dl\ is the distance between the
furthest two fixed positions. In the examples above, we get $\dl =
1$ for \texttt{**11*} and \texttt{***10}, and $\dl = 4$ for
\texttt{1*1*0}.
\item The \emph{order} $o$ of a schema is the number of fixed positions it
contains. In the above example $o$ is 2, 2 and 3 respectively.
\end{itemize}

With these measures and $L$ as the total length of a chromosome, we
can now write
\beq\label{GAfund2}
n_S(t+1) \geq n_S(t) \cdot \frac{\bar{f}(S)}{\bar{f}_\mathit{total}}
\left[ 1 - \frac{\delta(S)}{L-1} - o(S)\cdot p_m \right].
\eeq
The first correction term in the square brackets includes the effect
of crossover on the schema we are counting. With a probability of
$\frac{\dl(S)}{L-1}$, the crossover site lies within the schema and
the schema
may get destroyed. Of course, some crossovers will preserve the
schema even in that case, namely when by chance the partner in the crossover
provides the right bits in the right positions. Therefore equation
\ref{GAfund2} 
only gives a lower bound for the number of schema $S$ in the new
generation.

The final term is the effect of mutation on a schema. In a schema of
order $o$, there is a probability of $(1 - p_{mut})^o$ that the schema
survives mutation. For small $p_{mut}$, as is usually the case, one
can write $(1 - p_{mut})^o \approx (1 - o \cdot p_{mut})$.

A consequence of equation \ref{GAfund2} is that short, low-order
schemata of high fitness are \emph{building blocks} toward a solution
of the problem. During a run of the GA, the selection operator ensures
that building blocks associated with fitter individuals propagate
throughout the population.
The crossover operator ensures
that with time, several different good building blocks come together in one
individual to bring it closer to the optimal solution.
One can show that in a population of size $N$, approximately
$\mathcal{O}(N^3)$ schemata are processed in each generation \cite{goldberg}.

\subsection{Advanced Operators}\label{encoding}
The basic GA we have just introduced can be extended in many ways to
address specific 
problems. Variations are possible at almost any step, but we describe here the
variations that we found useful in solving our function maximisation problem.
The approach that worked best was a modification of the Breeder
Genetic Algorithm presented in \cite{breeder}. It uses a real valued
encoding of the problem parameters rather than the binary encoding
presented before. This also requires an adapted set of selection,
crossover and mutation operators. The following will summarise the choice of
operators that worked best for our problem.
\begin{description}
\item[Encoding:]
One chromosome consists of eight real numbers, directly representing the input
parameters (these are disscussed later, see table
\ref{input} on page \pageref{input}). Therefore, no decoding step is
neccessary: the phenotype 
is directly equivalent to the chromosome.
\item[Selection:]
Instead of the stochastic roulette selection mentioned earlier, which
can be seen as a model of 
natural selection,
we use \emph{truncation selection} which models the way a human
breeder might select 
promising candidates for mating.
 All individuals in a generation are sorted according to
their fitness and only the top third of individuals is taken to form the
breeding pool. There they are paired up randomly and offspring is
produced through crossover (see below) until a child population of
equal size to the parent population has
been created. Self-mating in the breeding pool is not permitted.
To prevent a degradation in the maximal fitness already achieved,
the best individual of the parent generation is copied into the child
generation unchanged.
\item[Crossover:]
Crossover is implemented as \emph{intermediate recombination}.
Take the chromosomes of both parents to represent two points $A$ and
$B$ in their eight-dimensional parameter space. Now imagine a
hypercube aligned with the coordinate axes,
with $A$ and $B$ at the endpoints of the longest diagonal.
The child's chromosome will then be picked at random from within this
hypercube\footnote{Actually, the child chromosome is picked from a
  hypercube which is larger by 25\% along each direction than the one
  spanned by $A$ and $B$, to prevent a rapid contraction of the search
  space towards values that lie centrally.}. 
\item[Mutation:]
After creation of one child chromosome by crossover, 
the mutation operator is applied to each one of the eight parameters
in the chromosome with a probability of $0.25$. 
If a value $x$ is to be mutated, a shift $\delta x$ is either added or
subtracted from $x$ with equal probability.
$\delta x$ is determined anew every time it is used through
\beq
\delta x=R\cdot\sum_{i=1}^{20}\left(2^{-i}\cdot P_{0.05} \right)\,;\qquad
P_{0.05}=\left\{ {1 \mbox{ with probability } 0.05} 
\atop {0 \mbox{ with probability }
  0.95} \right.\,,
\eeq
where $R$ is the range from the smallest permitted value for $x$ to the
largest permitted one. It is possible for the mutated $x$ to lie
outside the permitted 
range. If this happens, $x$ is reset to the minimum or maximum allowed
value respectively. Note that the definition for $\delta x$ creates
small perturbations much more often than large ones\footnote{This also
explains the rather high value of 0.25 for the overall mutation rate. 
If one only takes mutations above a certain magnitude to be
significant in changing the characteristics of the individual, the
occurence rate of such mutations  
would be much closer to the values 0.001--0.01 mentioned
before.}. This leads to 
a good search behaviour in finding an optimum locally, but also
to a good coverage of the full parameter space.
\end{description}
We found this set of operators to give the best convergence behaviour
for our problem. To completely state all GA related data here, we have
run the algorithm with a population size of 300. The runs were stopped
when no more improvement in maximal fitness happened for the 
 last 20 generations.

\section{Explicit Examples}
We choose  the models discussed in Refs.~\cite{agq,aaikq} as candidates to
discriminate. This choice is arbitrary, intended just to exemplify the
technique, which should apply in principle to any models. We note in passing
that using the masses is also an arbitrary choice, and in principle one could
choose any observables as the dimensions of space $\mc{M}$.
For concreteness we are using models that have been studied in
type I string theory in which the source of supersymmetry breaking
are either the dilaton field $S$, the overall size of the internal
manifold $T$ and a blowing-up mode $B$. A combination of their 
$F$-terms break supersymmetry and they can be parametrised by two
goldstino angles:
\beq
\bmat{c}F^S\\F^T\\F^B\emat=\bmat{c}\sin\theta\\
\cos\theta\sin\phi\\\cos\theta\cos\phi\emat
F_{total},
\eeq
where $F_{total}=\sqrt{(F^S)^2+(F^T)^2+(F^B)^2}$.

\subsection{The scenarios}\label{scenarios}
In this section we want to summarise the three scenarios that will be
used in the remainder of the work. First,  the soft breaking terms
for all three are:
\beq\label{m0used}
m_0^2=\mgr^2\left(1-
           \cos^2\theta\sin^2\phi\right)\,,
\eeq
\beq\label{m12used}
M_a=\sqrt3\mgr{\alpha_a\over\agut}\left(\sin\theta
      -{\beta_a\over4}\agut\,\cos\theta\left(
            {-10\over\sqrt{3}}\sin\phi + \cos\phi\right)
      \right)\, ,
\eeq
\beq\label{Atermused}
A_{\alpha\beta\gamma}= -\sqrt3
\mgr\sin\theta\,.
\eeq
$m_0$ is a flavour-family universal scalar mass, $M_{a=1,2,3}$ the 
mass of the U(1), SU(2)$_L$ and SU(3) gauginos respectively.
Defining a Yukawa coupling $Y_{\alpha \beta \gamma}$, 
$A_{\alpha \beta \gamma} Y_{\alpha \beta \gamma}$ (no summation on repeated
indices) is the
trilinear interaction between scalars denoted by $\alpha \beta \gamma$.
$\beta_a$ is the RGE $\beta$ function of gauge group $a$.
As free parameters in the SUSY breaking sector,
we have now \thet, \ph\ and \mgr. Additionally,
\tanb\ and the sign of $\mu$ are free parameters which are chosen in order to
fix $m_3^2$ and $|\mu|$ (in the notation of Ref.~\cite{softsusy}) from the
minimisation of the MSSM Higgs potential and $M_Z$. 

Note that the scalar masses and the trilinear $A$-terms are universal
and quite straightforward. Their values at the high scale also do not
depend on the choice of gauge unification behaviour. The gaugino
masses, however, do depend on the different gauge running behaviours, 
as one would expect, through their dependence on
$\alpha_a$ and $\alpha_{GUT}$. This
leads us 
to the three model scenarios and their abbreviations which 
will be frequently referred to from now on:
\begin{itemize}
\item The \fbox{\textbf{GUT}} scenario, with the string scale at
  $M_S=2\times10^{16}\,\gev$ and the standard MSSM particle content. In
  this scenario we
  chose the usual unified gauge coupling 
  $\alpha_a=\alpha_{GUT}=1/25$ as input for the soft terms.
  This will not make the soft gaugino masses universal, as they
  still contain the dependence on $\beta_a$.
\item The early unification scenario \fbox{\textbf{EUF}}, where
  $M_S=5\times10^{11}\,\gev$. To make unification work at this scale, we
  have added $2\times L_L+3\times e_R$ vector-like representations to
  the MSSM particle spectrum. We assume that their Yukawa couplings
  are negligible. Their effects are set to modify the one-loop
  beta function coefficients $\beta_1, \beta_2$ above a scale of $1\tev$.
This 
  happens to be the 
  simplest possible additional matter content that achieves the
  desired effect. There are models that contain possibly suitable
  candidates for such extra fields. Here $\alpha_a=\alpha_{GUT}=1/21$.
\item The mirage scenario \fbox{\textbf{MIR}} \cite{mirage}. The
  fundamental string scale 
  is again $M_S=5\times10^{11}\,\gev$, but now the gauge couplings are
  set independently to $\alpha_1=1/37.6$, $\alpha_2=1/27$ and
  $\alpha_3=1/19.8$, while 
  $\alpha_{GUT}=1/25$ remains at the usual GUT value. 
\end{itemize}
To predict sparticle masses from these scenarios, we must solve the
renormalisation group equations (RGEs)
starting from a theoretical boundary condition parameterised by 
the string scale $M_S$, the goldstino angles $\theta$ and $\phi$, 
the ratio of Higgs VEVs $\tan \beta$, and the gravitino mass $m_{3/2}$. 
Constraints from experiments and cosmology (if a version of $R$-parity is
conserved, as assumed here) restrict the models further. 

\subsection{Constraints}
We use the following experimental constraints to 
limit the scenarios~\cite{pdg,susywg}:
\begin{equation}
m_{\tilde\chi^0_1}>45\,\gev\qquad 
m_{\tilde\chi_1^{\pm}}>103\,\gev\qquad 
m_{h_0}>113.5\,\gev, \label{constr}
\end{equation}
\begin{equation}
-4.2<\delta a_\mu \times 10^{10}< 41.3\,.
\end{equation}
Also, the neutralino must be the LSP. 
Any parameter choice violating one or more of these constraints is considered
to be outside the footprint.
Throughout the whole analysis $\mu>0$ and $m_t=175\gev$ were
assumed~\cite{pdg}. 
Negative $\mu$ leads to a negative $\delta a_\mu$ SUSY contribution, which is
limited from 
the measurement of $(g-2)_\mu$~\cite{bnl,narison} to be small in magnitude. 
This means that, for a given
value of $\tan \beta$, the sparticles must be heavy in order to suppress their
contribution to $\delta a_\mu$. In this limit, effects of the sign of $\mu$
upon the mass spectrum are suppressed. We can therefore safely ignore the 
$\mu<0$ case because its resulting spectra will be included in our $\mu>0$
results.

Table~\ref{input} shows the default ranges allowed for the parameters. It 
also summarises the other parameters that were kept constant. $m_{3/2}$ is
restricted not to be too big, since then one introduces too much fine tuning
in the Higgs sector of the MSSM~\cite{fineTuning}, $\tan \beta$ is bounded
from above by the constraint of perturbativity of Yukawa couplings up to $M_i$
or $M_{GUT}$, and from below by LEP2 Higgs
data~\cite{susywg}. $\theta>30\dg$ is chosen to avoid a situation where 
anomaly-mediated SUSY breaking effects are comparable to gravity mediated
ones, for which the pattern of soft SUSY breaking terms is currently unknown.
\TABULAR[t]{|c|c|c|c|c|c|c|c|}
{\hline\label{input}
$\theta$ & $\phi$ & $m_{3/2}$ & $\tan \beta$ & $\mu$ & $m_t$ &
  $M_{GUT}$ 
& $M_I$\\
30--90$\dg$& 0--90$\dg$ & 50--1500 & 2--50 & $>0$ 
& 175 
& $2\times 10^{16}$ &$5\times 10^{11}$ \\
\hline
}
{Summary of parameters. The first four parameters are scanned over, and their
range is given. The value of the others is kept constant except for
$\mu$ which is constrained to give the 
correct value of $M_Z$. 
All massive parameters ($m_{3/2}, m_t, M_I, M_{GUT}$) are given in units of
GeV.} 

\section{Results}
As maximisation criterion for the genetic algorithm 
we use the inverse of the relative distance $\Delta$, which we defined
earlier (\ref{dist-measure}). In our scenarios, this quantity
 is built from the sparticle masses at two model points $A$ and $B$ as
 follows:  
\beq
\mathit{Fitness}\equiv
\frac{1}{\Delta} = \frac{|\vec{M}_A+\vec{M}_B|}{|\vec{M}_A-\vec{M}_B|}
=\sqrt{\frac{%
\bigg(M_{\susy{\chi}^0_1,\,A} + M_{\susy{\chi}^0_1,\,B}\bigg)^2
+\ldots
+\bigg(M_{\susy{\tau}^{}_2,\,A} + M_{\susy{\tau}^{}_2,\,B}\bigg)^2}{%
\bigg(M_{\susy{\chi}^0_1,\,A} - M_{\susy{\chi}^0_1,\,B}\bigg)^2
+\ldots
+\bigg(M_{\susy{\tau}^{}_2,\,A} - M_{\susy{\tau}^{}_2,\,B}\bigg)^2}}\;.
\eeq
The full list of masses we used can be found in table
\ref{tab:spectra}.

The GA is run
until no improvement is seen for 20 generations in a row, with
populations of 300 individuals. We initially found that runs using binary
coding were unstable: successive runs with different random initial conditions
gave significantly different fitnesses. We therefore switched to the real
encoding mentioned in section~\ref{encoding}, which we find to work. 
An important criterion is that the method can tell when two models are
truly non-distinguishable, \ie their footprints overlap: We should see large
fitness values of the order of the inverse numerical precision of the
calculation. 

\FIGURE[]{
\fourgraphs{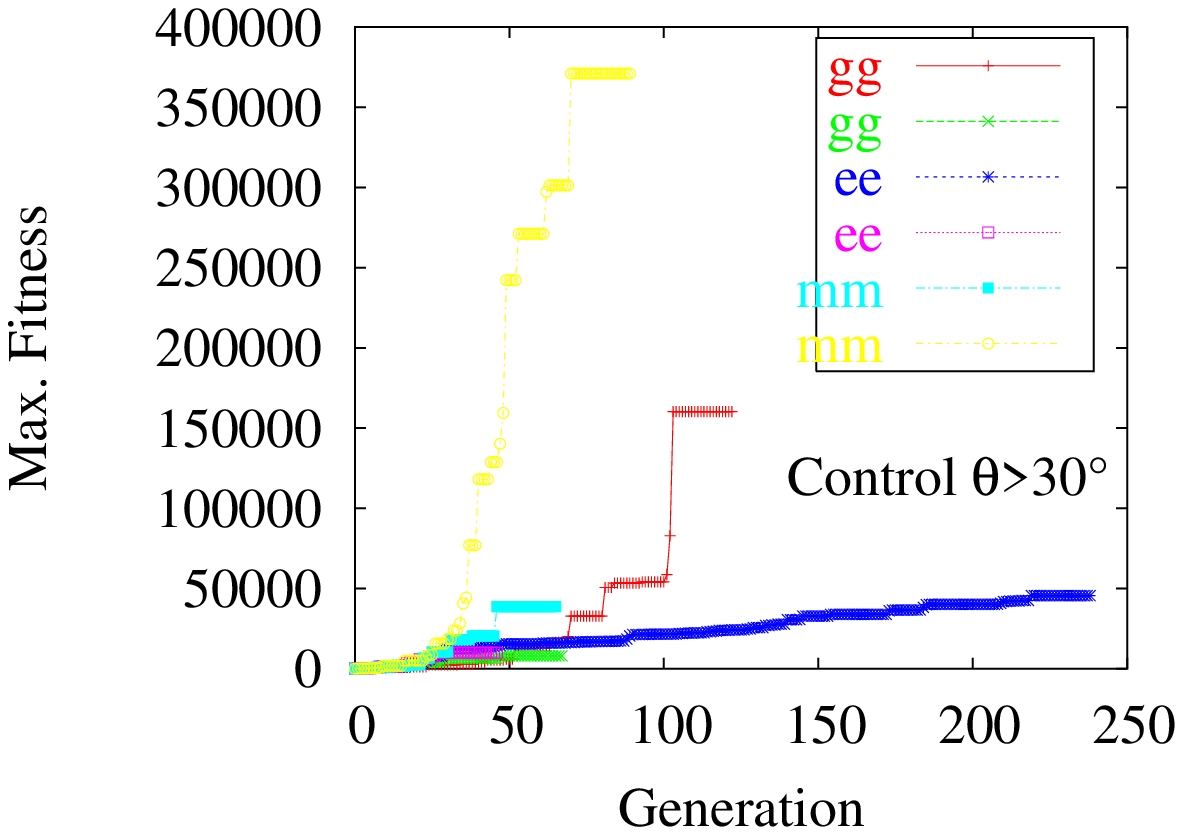}{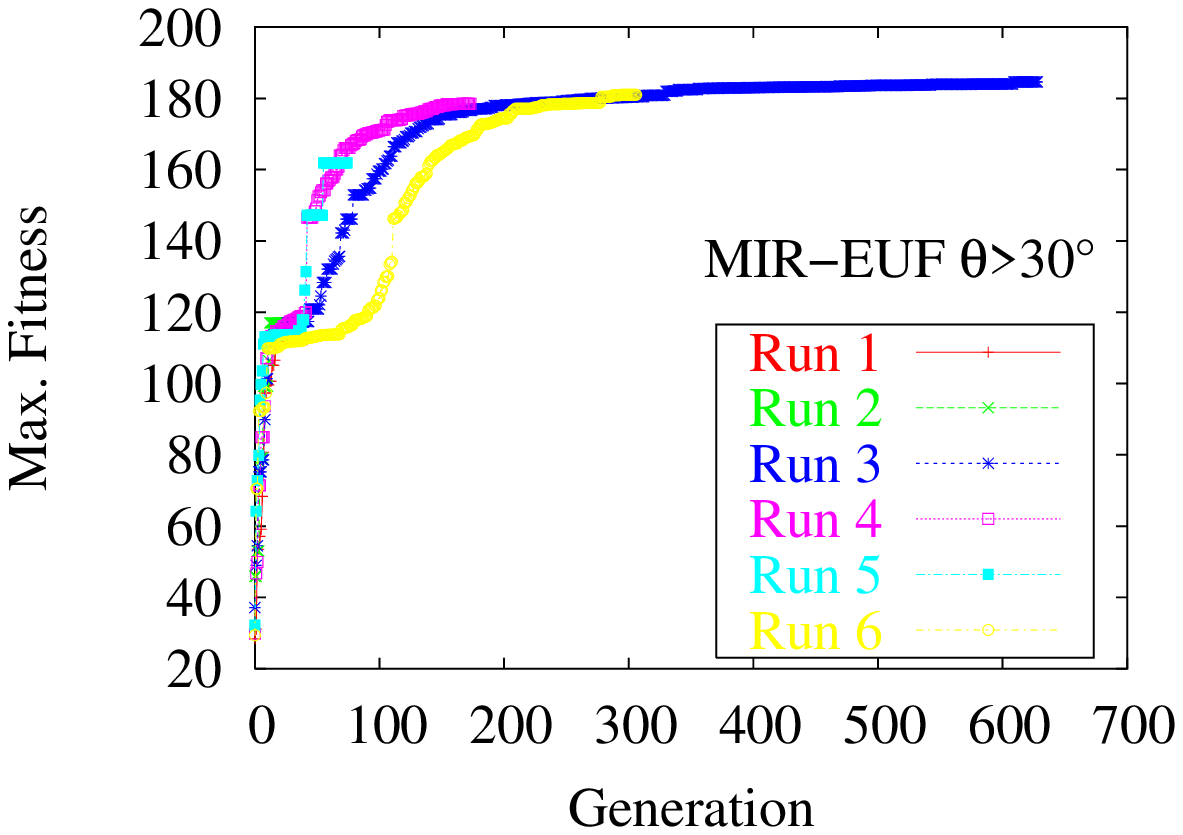}{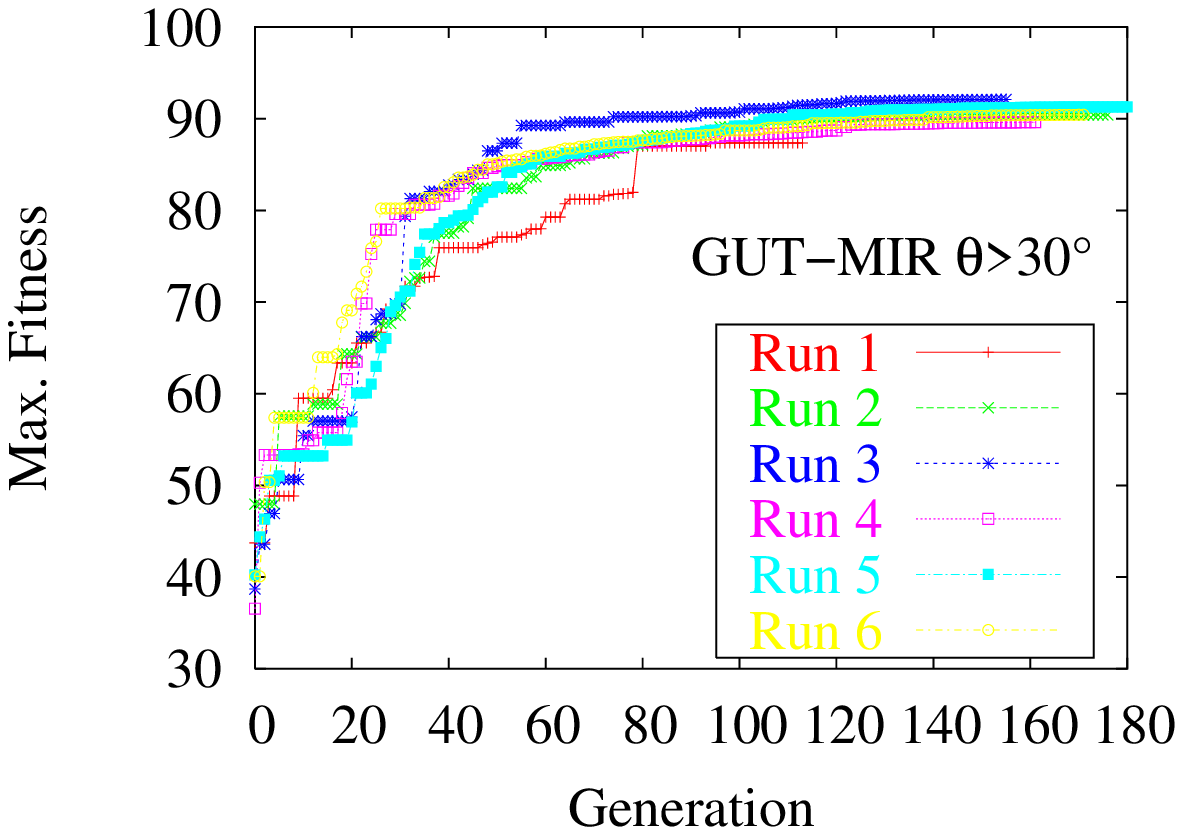}{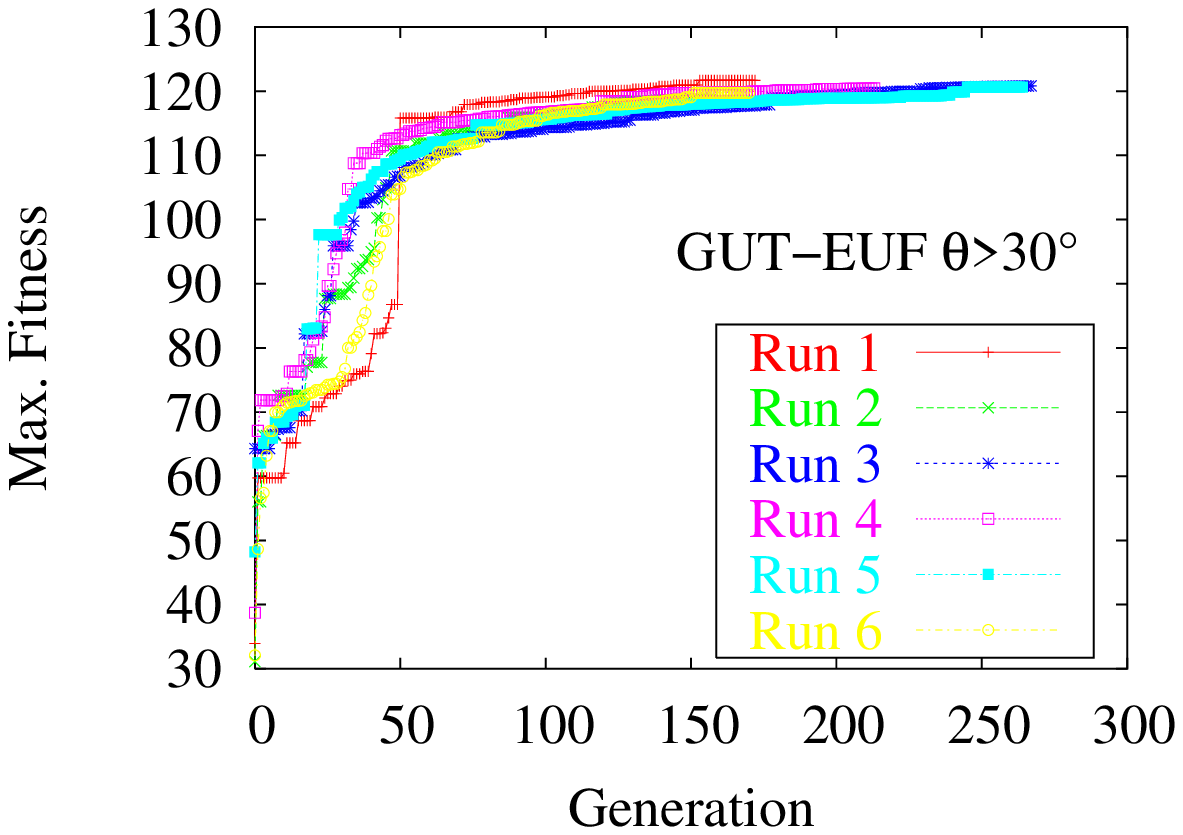}
\caption{GA Progress in model discrimination: (a) control samples for GUT-GUT
  (gg), early unification - EUF (ee) and mirage-mirage (mm)
  ``discrimination'', 
(b) mirage-early unification, (c) GUT-mirage, (d) GUT-early unification.
The evolution of the best individual's fitness with generation is plotted.}
\label{GAprogress}
}
Fig.~\ref{GAprogress}a illustrates that this is indeed is the case
when we choose 
the two models to be {\em identical}. The numerical
precision of the SOFTSUSY calculation was set to $10^{-5}$, 
 and we obtain fitnesses of order the inverse of this
number in each case. Three separate runs were tried for the GUT scenario
(gg), two for the early unification (ee) and two for the mirage
unification scenarios (mm). Each run was started with independent random
numbers. We see that in each case a large fitness, $\mc{O}(10^5)$ results.
The progress of the discrimination runs show a different
pattern, as shown in Figs.~\ref{GAprogress}b--d for mirage-early unification,
GUT-mirage and GUT-early unification discrimination respectively. In each
case, six independent 
runs are tried and in each case we see that the fitness converges to a stable
value after 100--500 generations. The rate of progress in early
generations varies 
depending upon the random numbers used to seed the algorithm, but each
separate run converges to the same value. 
The fitnesses in the discrimination runs are much lower than in the control run
(Fig.~\ref{GAprogress}a), indicating that the three footprints in $\mc{M}$ are
disjoint. As an example, we show the input parameters for the best
pairs in the 6 different runs
for the GUT-EUF discrimination (Table~\ref{tab:inputs}). 
{\small
\TABULAR{|c|cccccc|cccccc|}{
\hline
Model & \multicolumn{6}{c|}{GUT} & \multicolumn{6}{c|}{EUF} \\ \hline
Run  &1 & 2 & 3 & 4 & 5 & 6 &1 & 2 & 3 & 4 & 5 & 6 \\ \hline
\thet&	51 & 49 & 49 & 50 & 51 & 49 & 85 & 77 & 76 & 81 & 85 & 85 \\	 
\ph&	35 & 40 & 40 & 41 & 39 & 28 & 51 & 34 & 31 & 49 & 58 & 25 \\	 
\tanb&	4.8 & 5.1 & 3.8 & 4.8 & 4.5 & 8.0 & 3.3 & 3.4 & 2.9 & 3.3 & 3.2 & 4.2
	 \\
\mgr& 0.85 & 1.10 & 1.14 & 1.02 & 0.99 & 0.97 & 0.82 & 1.05 & 1.09 & 0.97 &
0.96 & 0.93 \\ \hline		 
Max.~fitness & {122} &{119} &{121}
&{120} &{121} &{109}& {122} &{119} &{121}
&{120} &{121} &{109}\\ \hline
}{Input parameters of closest points (smallest $\Delta$) in
  6 independent GUT-early unification     discrimination runs. The
  angles $\phi,  \theta$ are  
  listed in degrees and \mgr\ is listed in TeV. \label{tab:inputs}
}}

Although the fitnesses of the solutions are similar in independent runs
(within one plot), the
actual positions in the input parameters (and the observables) are
different. This occurs because of approximately degenerate minima, whenever 
the boundaries of two footprints in $\mc{M}$ are roughly parallel in
some region of the parameter space. 
\TABULAR{|c|cc|cc|cc|cc|}{
\hline
&		MIR&	   EUF&      MIR&     EUF&     MIR   & GUT   &GUT   &
EUF  \\ \hline
\thet&		 42.2 &77.2 &45.4 &87.5 &51.2 &66.2 &51.1 &85.3 \\
\ph&		  33.2 & 36.3 & 24.4 & 0.2 & 33.2 & 57.9& 35.4 & 51.1\\
\tanb&		  3.4 & 4.2 & 3.3 & 3.9 & 3.1& 6.1 & 4.8 & 3.3\\
\mgr&		  1194 & 991 & 1080 & 937 & 979 & 810 & 845 & 820\\
\hline
$\neut{1}$&	  673 & 657 & 603 & 585 & 613 & 619 & 551 & 532\\
 $\neut{2}$&	  1158 & 1156 & 1099 & 1095 & 1093 & 1062 & 941 & 962\\
$\neut{3}$&	  1639 & 1652 & 1642 & 1658 & 1632 & 1685 & 1536 & 1499\\
$\neut{4}$&	  1650 & 1661 & 1652 & 1666 & 1642 & 1692 & 1543 & 1509\\
$\charg{1}$&	  1158 & 1156 & 1098 & 1095 & 1093 & 1062 & 941 & 962\\
$\charg{2}$&	  1649 & 1661 & 1651 & 1665 & 1641 & 1691 & 1542 & 1508\\
$h^0$&		  115 & 119 & 114 & 118 & 113 & 121 & 119 & 115\\
$A^0$&		  2191 & 2178 & 2158 & 2151 & 2101 & 2067 & 1932 & 1951\\
$H^0$&		  2192 & 2178 & 2159 & 2151 & 2102 & 2067 & 1933 & 1952\\
$H^\pm$&	  2193 & 2180 & 2160 & 2153 & 2103 & 2069 & 1934 & 1953\\
$\s{g}$&	  2851 & 2873 & 2830 & 2847 & 2770 & 2646 & 2362 & 2477\\
$\s{u}_L$&	  2609 & 2631 & 2564 & 2585 & 2482 & 2526 & 2287 & 2260\\
$\s{d}_L$&	  2610 & 2632 & 2565 & 2586 & 2483 & 2528 & 2288 & 2261\\
$\s{u}_R$&	  2524 & 2512 & 2485 & 2474 & 2401 & 2408 & 2184 & 2161\\
$\s{d}_R$&	  2516 & 2493 & 2479 & 2459 & 2395 & 2387 & 2166 & 2147\\
$\s{t}_1$&	  1916 & 1859 & 1898 & 1841 & 1834 & 1752 & 1571 & 1589\\
$\s{t}_2$&	  2363 & 2369 & 2328 & 2332 & 2258 & 2261 & 2045 & 2041\\
$\s{b}_1$&	  2505 & 2482 & 2468 & 2448 & 2385 & 2373 & 2155 & 2138\\
$\s{b}_2$&	  2335 & 2339 & 2298 & 2302 & 2227 & 2234 & 2017 & 2008\\
$\s{\nu_e}$&	  1302 & 1324 & 1233 & 1257 & 1139 & 1164 & 1110 & 1106\\
$\s{e}_L$& 	  1304 & 1326 & 1235 & 1259 & 1141 & 1166 & 1113 & 1109\\
$\s{e}_R$&	  1148 & 1122 & 1082 & 1055 & 976 & 941 & 929 & 931\\
$\s{\nu_{\tau}}$&  1302 & 1323 & 1232 & 1256 & 1138 & 1161 & 1109 & 1106\\
$\s{\tau}_1$&	  1146 & 1119 & 1081 & 1053 & 974 & 934 & 925 & 929\\
$\s{\tau}_2$&	  1304 & 1325 & 1234 & 1258 & 1140 & 1164 & 1111 & 1108\\
\hline
$\Delta$ &\multicolumn{2}{c|}{0.0054}&\multicolumn{2}{c|}{0.0055} 
&\multicolumn{2}{c|}{0.0109}&\multicolumn{2}{c|}{0.0082}\\
Fitness &\multicolumn{2}{c|}{184.6}&\multicolumn{2}{c|}{181.0} 
&\multicolumn{2}{c|}{92.1}&\multicolumn{2}{c|}{121.7}\\
\hline}{MSSM Spectra of closest-fit points in the various comparisons. The
  mirage (MIR) and early unification (EUF) scenarios are shown twice, for two
  different (approximately degenerate) maximal fitness. We also show the
  closest spectra from the GUT-mirage comparison and GUT-early unification
  comparison. All masses are in units of GeV. The angles $\phi, \theta$ are
  listed in degrees. \label{tab:spectra}}

We give examples of the observables (in this case, masses) corresponding to
the closest-fit points of each footprint in Table~\ref{tab:spectra}. For
comparison purposes, two examples are picked from MIR-EUF discrimination
runs in order to show how much the MSSM spectra differ at two different
closest-fit points. The spectra are similar, as is evident by comparing the
columns 2 and 4 of Table~\ref{tab:spectra}, or columns 3 and 5. Again, 
this does not have to necessarily be the case (but proved to be the case in
our results). We see that our intuition that $\Delta$ roughly measures  the
sort of fractional precision one needs to measure the observables, in order to
discriminate two models, holds by comparing the spectra between the two models
that are being discriminated. We also see that accurate $\neut{1}$, $h^0$,
$\s{t}_1$, $\s{\tau}_1$ mass measurements are important to help
discriminate between 
the MIR and EUF scenarios (since these show larger mass differences).

\section{Conclusions}
Genetic algorithms have allowed us to 
answer the problem of discrimination of SUSY breaking models in the
following questions: what accuracy on measurements is required to reliably tell
two given 
different SUSY breaking scenarios apart, and which are the most important
variables to measure? 
We have studied the discrimination of
three different SUSY breaking scenarios as examples, and
assumed that the relevant observables are the masses. Each of these
assumptions is arbitrary, and the GAs can be applied in other situations
where one wants to discriminate different models using different observables.
More standard approaches such as scans or hill-climbing algorithms did not
yield stable solutions. 

We have constructed a measure of ``relative distance'' that describes the
relative difference between two MSSM mass spectra. This in principle uses the
entire spectrum rather than some subset~\cite{agq} in order to parameterise
discrimination. We found that each model can be in principle discriminated
from the others, in a total of 3 comparisons. 
Values corresponding to $\Delta$ of 
0.5$\%$, 1$\%$ and 1$\%$ were found in the three
comparisons, indicating that this is the rough accuracy that will be required
for sparticle mass measurements and predictions~\cite{predErrors} 
in order to distinguish the models. In a
control sample of a model to be discriminated against 
itself, a fractional accuracy of  $\Delta=0.001\%$ is found, corresponding to
the numerical accuracy of the calculation. This indicates that the two
scenarios are indeed indistinguishable, providing confidence in the method.
For more precise information regarding the simultaneous accuracies
that are required for discrimination, the spectra predicted by the two points
with smallest relative distance must be compared. This information is
difficult to use, and would become more relevant when one knows which
dimensions of $\mc{M}$ to use (corresponding to the minimal set of
measurements that need to be made). 
In fact, the point pairs found to have the smallest ``relative
distance'' vary in successive runs, indicating some approximately degenerate
minima.

Now that a working setup of the GA minimisation procedure has been found, many
possible applications beyond the test scenario introduced here could
be taken into consideration:
\begin{itemize}
\item First of all, before experimental data becomes available, we can
  compare more model footprints in exactly the way described above,
  and can try to find classes of models 
  that should easily be distinguishable from others.  This is of
  course not restricted to models motivated by string theory, but can
  be applied to any kind of model that makes predictions about the
  low-energy sparticle mass spectrum. Possibilities include different
  sets of gauge mediated SUSY breaking scenarios, or a comparison with
  models where SUSY breaking is purely anomaly mediated.
\item One could also take some typical test scenarios of different
  models (rather 
  than the closest ones) and test discrimination power based on what
  measurements various future colliders are expected to deliver. 
\item As soon as real sparticle measurements are available, the GAs
  could take on a new role. Such an actual measurement will pick out a
  hyperplane, a line or even a point in the measurement space (depending on how
  many of the masses have been determined). 
  The experimental
  uncertainty will blur out these objects somewhat, leading to
  something quite like the footprints in the preceding chapters. 
  Minimising $\chi^2$ in the MSSM for assumed SUSY breaking models has been 
  addressed using a combination of scan and hill-climbing
  algorithms~\cite{sfitter}. However, we suspect that GAs may provide a more
  robust solution for finding a $\chi^2$ minimum. 
\item The GA approach also
  leads to an elegant way of dealing with the 
  problem of fine-tuning in this last proposal, since one could now define the
  footprint to contain the experimentally acceptable solutions and minimise
  the fine tuning within them.  
\end{itemize}
These techniques may also 
 be useful in other areas. We can
mention at least two of them. First, 
in cosmology, due to the level of precision that the observations are
achieving, particularly for the cosmic microwave background (CMB) and
due to the success of inflation to explain the current observations,
we are in  a situation similar to the one considered here because, as in
the case of supersymmetric models,
 there are plenty of models of cosmological  inflation.
An important task for the future is to find efficient ways to
discriminate among different models of inflation. More generally, the
 parameter set having a better fit with data needs to be investigated
 systematically (see \cite{liddle} for an interesting
 discussion in this direction).  Genetic algorithms may be of
 use in this effort.

Secondly, in string theory. There is an increasing evidence that the
number of string vacua is huge. Statistical techniques are actually
starting to be used in order to study classes of vacua \cite{douglas}
and genetic algorithms may play an important role in this effort. 
In particular, 
there is an outstanding problem of how to discriminate among 
different compactifications and we may find genetic algorithms useful
in a similar way in which we have applied them here.

Genetic algorithms have shown their
robustness and power in many other widely separate fields of
engineering and research, and there is no reason why theoretical 
particle physics should be an exception. 

\section*{Acknowledgements}
We thank S.~Abel, A.~Davis, H.~Dreiner,  D.~Ghilencea, G.~Kane and 
H.P.~Nilles for
interesting conversations. D.G. would especially like to thank F.~Dolan
for the discussions on the distance measure. 
The research of F.Q. is partially supported
by PPARC and the Royal Society Wolfson award. Parts of this work were
supported by the European Commission RTN programs HPRN-CT-2000-00131,
00148 and 00152.

\end{document}